# Dynamic FPGA Detection and Protection of Hardware Trojan: A Comparative Analysis

**Amr Alanwar[1,2], Mona A. Aboelnaga[1], Yousra Alkabani[1,3], M. Watheq El-Kharashi[1,4*] and Hassan Bedour[1]**

[1]*Computer and Systems Engineering Department, Ain Shams University, Cairo, Egypt*
[2]*Electrical Engineering Department, University of California, Los Angeles, Los Angeles, CA, USA*
[3]*Computer Science and Engineering Department, American University in Cairo, Cairo, Egypt*
[4]*Electrical and Computer Engineering Department, University of Victoria, Victoria, Canada*

**Abstract:** Hardware Trojan detection and protection is becoming more crucial as more untrusted third parties manufacture many parts of critical systems nowadays. The most common way to detect hardware Trojans is comparing the untrusted design with a golden (trusted) one. However, third party intellectual properties (IPs) are black boxes with no golden IPs to trust. So, previous attempts to detect hardware Trojans will not work with third party IPs. In this work, we present novel methods for Trojan protection and detection on field programmable gate arrays (FPGAs) without the need for golden chips. Presented methods work at runtime instead of test time. We provide a wide spectrum of Trojan detection and protection methods. While the simplest methods have low overhead and provide limited protection mechanisms, more sophisticated and costly techniques are introduced that can detect hardware Trojans and even clean up the system from infected IPs. Moreover, we study the cost of using the FPGA partial reconfiguration feature to get rid of infected IPs. In addition, we discuss the possibility to construct IP core certificate authority that maintains a centralized database of unsafe vendors and IPs. We show the practicality of the introduced schemes by implementing the different methodologies on FPGAs. Results show that simple methods present negligible overheads and as we try to increase security the delay and power overheads increase.

## 1. Introduction

There is a dramatic shift from the vertical integrated silicon industry business to the horizontal one due to the rising cost of preserving manufacturing facilities, the increase in the competition, and the desire to have shorter time to market. Most design houses are currently fabless and tend to manufacture their designs in offshore facilities. Furthermore, designs are now complex so designers tend to embed some parts of the design by third party IPs. However, using third party IPs in the design industry raises many trust issues. The details of third party IPs are hidden and the fabrication process is obscured from the designer to protect the IP owner's rights. Designers need to ensure that no malicious circuitry is embedded in their designs by third party IP owners.

With the widespread use of silicon chips in different applications, varying from cell phones, cars, to strategically important military devices; it is vital to provide methods that resolve the trust issues between fabrication facilities, designers, and end-users. One of the most important problems to resolve is hardware Trojan horse malware (malicious altering of hardware specification or implementation, denoted as Trojan thereafter) circuit protection and detection. Designers need to have guarantees that their designs are not tampered with while maintaining technology secrets of the fabrication facilities and design royalties of third party IP owners. End-users need to get guarantees that their devices are not controlled by unknown entities and/or will not leak sensitive information about them.

Hardware Trojan detection methods have been developed to ensure that no malicious circuitry is embedded in their designs. These methods either try to introduce architectural modifications to prevent embedding Trojan [1, 2, 3, 4], or try to detect the existence of a Trojan with a high probability by studying side-channel waveforms [5, 6, 7, 8, 9, 10]. However, these methods mainly depend on comparing the untrusted chip with a trusted golden one. In practice, there is no golden

*E-mail address: {amr.alanwar, mona_alaa, yousra.alkabani, watheq.elkarashi, hassan.bedour}@eng.asu.edu.eg*

chip for third party IPs. Attempts were introduced to depend on the system integrator's design specifications for comparisons [11].

This paper develops novel methods for Hardware Trojan protection and detection. We introduce four methods that provide different levels of protection. Fig. 1 shows the taxonomy of the introduced schemes for dynamic FPGA system protection and detection of Trojans. We can divide the presented work into methods for simple protection and methods for both detection and protection. Simple protection methods include: (1) simple blockage (SB) and (2) multiplexing reconfigurable variants' outputs (MRVO). The dynamic protection and detection methods are: (1) multiplexing reconfigurable IPs' outputs and cyclic redundancy check (CRC) Trojan detection schema (MCRC) and (2) multiple variants implementation (MV).

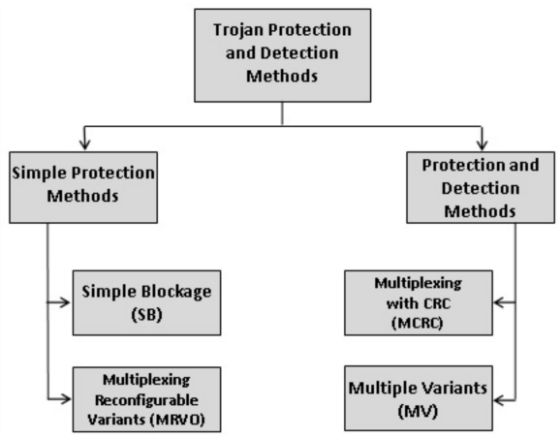

Figure 1: Taxonomy of FPGA Trojan detection and protection schemes.

The SB method works by obfuscating the output of suspected IPs and deobfuscating it only at the input of the receiving IPs. SB introduces low overhead. However, the protection it provides is limited. It can only protect from spying Trojans by hiding the information transmitted by the suspected IP. However, it will not really detect a Trojan or protect the circuit from failure. The MRVO method decreases the probability of circuit malfunction due to a Trojan attack by increasing the redundancy in the circuit by switching the operation of the circuit between at least two different implementations of the same IP. The MCRC method deals with infected Trojan by decreasing its probability to leak important information until detecting the infected IP and totally removing it by using the partial reconfiguration technique. The MV method introduces redundancy by using an odd number (at least three) of the same IP from different owners and voting according to the output.

The rest of the paper is organized as follows. Section 2 summarizes the related work. The protection against Trojan is described in Section 3, which consists of two subsections: 1) the implementation of the SB method and 2) the implementation of the MRVO method. Then, Trojan detection methodologies are described in Section 4, which consists of two subsections: 1) MCRC and 2) MV. We show CRC logger for safe use of third party IPs in Section 5. We introduce an authority to build a trusted IP market in Section 6. Experimental evaluation and estimation of the overhead of the methods are shown in Section 8. Finally, Section 9 concludes our work and presents future work directions.

## 2. RELATED WORK

Hardware Trojans are the malicious changing of hardware specification or implementation in a way that affects its functionality. Hardware Trojans are more dangerous than software Trojans because they reside at the lowest level of information processing and will continue to threaten the system as long as the infected hardware is in use.

One way to ensure that the presence of Hardware Trojans cannot affect a design is by preventing them from subverting the functionality of an IC. The ability to create trusted circuits is proposed by Potkonjak [12]. Proposed solution fully accounts for the use of all hardware resources at all times. Waksman and Sethumadhavan introduced a method that attempts to prevent Trojan triggering [13]. Beaumont *et al.* ran replica of a program on multiple processing elements to achieve protection from hardware Trojans [14]. Baumgarten *et al.* used reconfigurable logic barriers within a design to prevent the activation and operation of hardware Trojans added during the IC manufacturing stage [15]. Ibn Ziad *et al.* provided a protection technique and showed its overhead against the injection of hardware Trojans in a voting machine to tamper voting results [16]. The attack depends mainly on the unused bits. The power overhead was negligible while the delay overhead did not exceed 10%.

Hardware Trojan circuits are usually designed to be activated using a rare trigger. As a result, regular testing methods are not suitable to detect Trojans because they are not expected to be activated during testing. Recently, different methods have been specifically designed considering Trojan detection. We can classify these hardware Trojan detection methodologies into: (1) side-channel dependent methodologies and (2) architectural methodologies [17].

Trojan impact is localized by side channel-dependent methodologies. The main idea is to try to detect the presence of a Trojan with a high probability by detecting the overload of the malicious circuit on different circuit parameters; such as the critical path delay or the power consumed as compared to a Trojan-free circuit. Wang *et al.* depend on localized current analysis for Trojan detection. They analyze power from multiple ports to detect the impact of the existence of Trojan on power [18]. Rad *et al.* studied the impact of a Trojan on the power supply transient current of an IC using statistical methods [2]. Jin and Makris used critical path delay analysis to detect Trojans [8]. Banga *et al.* introduced a test vector generation method that can be used to differentiate between the side-channel waveforms of a Trojan-free and Trojan-inserted circuit [5, 6]. Gate-level characterization techniques accompanied by statistical methods are also used to detect hardware Trojans. These methods depend on gate-level power or delay characterization, or both [9, 10, 19, 20].

Architectural methodologies try to increase the opportunities of the activation of a hardware Trojan during testing. Salmani *et al.* increase the Trojan activity by inserting dummy flip flops in the design [4]. They chose locations of inserted flip flops based on a transition probability threshold. Banga and Hsiao used voltage inversion at alternating levels of the circuit to increase the power consumption of a Trojan-inserted circuit [1]. Rajendran *et al.* introduced a methodology for securing all the gates of a design using ring oscillators [3]. They added extra logic to convert paths of the circuit into ring oscillators. The presence of a Trojan is detected by the changes it causes in the frequency of the ring oscillators.

All above methods have a critical problem. They require a golden chip. This problem can be degraded if the design does not contain third party IPs [21]. However, if a system designer integrates third party IPs in the design, these methods become impractical. Zhang and Tehranipoor tried to provide an alternative to using a golden design by using code coverage analysis, formal verification, and Automatic Test Pattern Generation (ATPG) methods to achieve high confidence in whether the circuit is Trojan-free or Trojan-inserted [11].

### 3. SIMPLE PROTECTION AGAINST TROJAN METHODS

In this section, we show methods to protect from embedded Trojans without trying to detect them. Sometimes, designers cannot afford the cost of detecting Trojans. Hence, preventing sensitive information leakage without detecting the embedded Trojans becomes a designer target. We will introduce two simple methods: SB of Trojan outputs and MRVO.

#### A. Simple Blockage (SB) of Trojan Outputs

In the SB of Trojan outputs method, we try to provide protection by obfuscating the outputs of the untrusted IP before sending out data then undoing that obfuscation at the input of the receiver as shown in Fig. 2. This methodology is used to avoid revealing secret data through off-chip transmission [22].

Adhoc lightweight obfuscation functions can be used to achieve the target with low overhead. The inverse of the function needs to be used to retrieve original data at the receiver design part (Fig. 2). We suggest a simple function which is mainly based on data Xoring and Xnoring. Fig. 3 and Fig. 4 describe our simple obfuscation function and its inverse, respectively. Processing will be done on every 2 bits by Xoring or Xnoring them and moving one bit value to another. As shown in Fig. 3, the original data contains bits A and B, which will be Xored to give C. Then, B will be shifted to A's place. Function inverse is shown in Fig. 4, where B and C are Xored to retrieve A. Then, B is shifted back to its original place. The same will be done for the third and forth bits, but with Xnoring. If original data size is odd, the last bit can be permuted with one of the other bits at the end. Section 8 will show the obfuscation function overhead in details. We propose also to partially reconfigure an FPGA to replace the confusion function with another one periodically. This continuous update of the obfuscation function should make it difficult for the attacker to reveal the data.

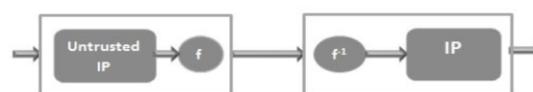

Figure 2: Using confusion circuits to protect a design against Trojan.

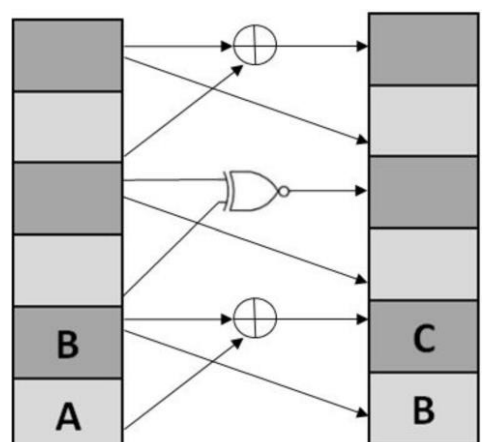

Figure 3: A sample of a confusion function to protect secret data.

#### B. Multiplexing Reconfigurable Variants Outputs (MRVO)

In the MRVO method, we try to increase the defense against Trojan attacks by reducing the probability of both leakage of sensitive information and malfunction due to a Trojan attack. We attempt to do this by using two or more untrusted implementations of the same IP from different untrusted vendors. Thus, the actual output of the system will be a mixture of the output from all different IP copies used in the system [23].

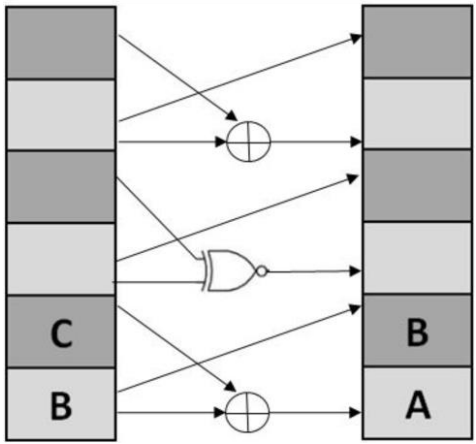

Figure 4: Inverse of the sample of the confusion function in Fig. 3 to be used to retrieve original data.

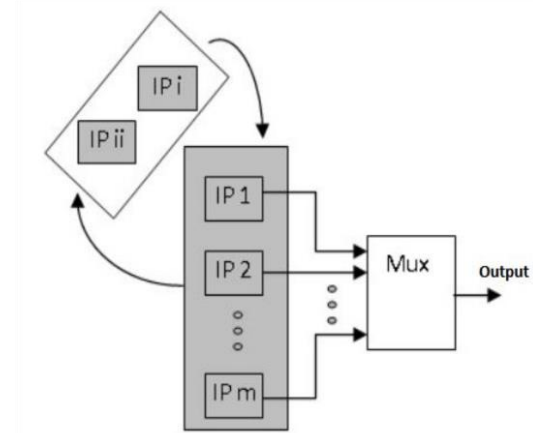

Figure 5: Multiplexing IP's outputs to reduce Trojan effect. IP1 to IPm can be replaced by IPi to IPii.

The main idea in case of not having a fully trusted IP is to use *m* copies of the IP from different untrusted vendors. At runtime, the output is randomly chosen from one of the *m* IPs. Thus, a Trojan will only impact the output if it is activated while the infected IP output is selected.

Furthermore, we suggest to partial reconfigure an FPGA to replace current running IPs periodically or even randomly to decrease the ability of system hacking. Fig. 5 illustrates the proposed idea. IP replacement takes about 8% of the time required to program the whole design on the FPGA, as will be explained in Section 8.

We propose the following criteria to select the output from different IPs.

*1) Unbiased Random Selection*

We use an unbiased random variable generator to select the output from one of the IPs. By this way, the probability of choosing an IP is equal among all the candidates. This methodology can be used if the design does not support the calculation of infection probability.

*2) Biased Random Selection*

We give a weighted voting among the available IPs that are used to choose output. Weights are inversely proportional to the probability of IP infection. Weights can be assigned using one of the following methods.

- Tracking IPs behaviors can be implemented to assign a suitable weight to each IP. The result of comparing the output of the running IPs on runtime can be used to assign weights. Also, the CRC of the IPs outputs can be used as a good candidate to build the learning curve and assign weights, as will be shown in Subsection 4.A.

- Weights can be assigned using the core *certificate, which is proposed in Section 6.*

## 4. TROJAN DETECTION METHODS

In this section, we show two Trojan detection methods: MCRC and MV and a scenario to remove infected IPs during runtime.

*A. Multiplexing Reconfigurable IPs' Outputs and CRC Trojan Detection Schema (MCRC)*

In this methodology, we multiplex *m* IPs output to select non infected-data. Our enhancement aims not only to decrease the probability of leaking sensitive information, but also to mark the IP as infected and its vendor as an attacker [24]. We introduce voting among IPs' CRCs to choose the major CRC, i.e., the CRC for the non infected-data. As shown in Fig. 6, the CRC voting circuit will be responsible for that job. Furthermore, the CRC voting circuit will keep track of the count of data errors for each IP producing infected output. If this count exceeds a certain threshold, the IP will be marked as an infected one and an alarm circuit will trigger a warning. We provide a small warning threshold for IPs that show discrepancies for the first time as this might be a design bug not a Trojan. However, this warning increases to include more actions when more faults are detected. The CRC voting circuit is shown in Fig. 7. It is worthy to mention that we can compare IPs outputs data instead of comparing their CRC. But, comparing IPs outputs will cost more area, power, and delay overheads as number of CRC output bits are much smaller in size than IPs'. This method will be very useful in building the core certificate authority, which will be in introduced in Section 6.

*1) Multiplexing Reconfigurable IPs'*

To further decrease the probability of information loss, we suggest using the FPGA partial reconfiguration (PR) feature which divides FPGA logic design into two different types: reconfigurable logic and static logic. Partial reconfiguration allows modifying reconfigurable regions in the FPGA without compromising the integrity of the applications running on the static logic part, as illustrated in Fig. 8. In our methodology each IP core reserves a reconfigurable logic region, whereas the dynamic Trojan detector unit and the Trojan alarm handling unit will reserve a static logic region.

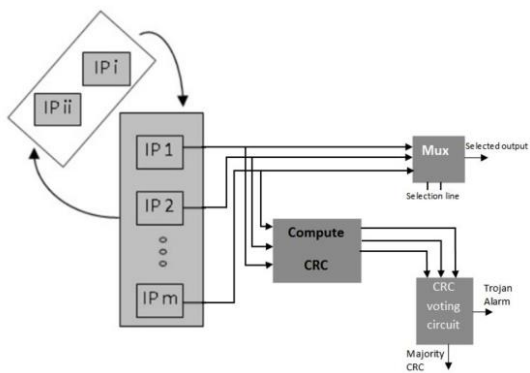

Figure 6: Multiplexing reconfigurable IPs' outputs to select a correct one and adding the CRC part to check output.

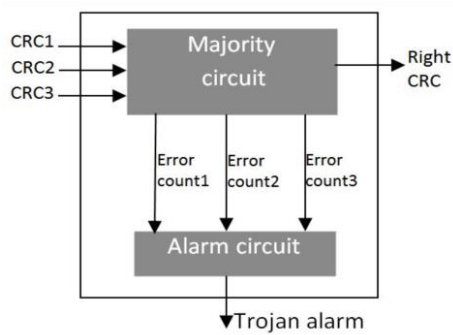

Figure 7: CRC voting circuit. Error count indicates number of times the CRC of each IP is faulty.

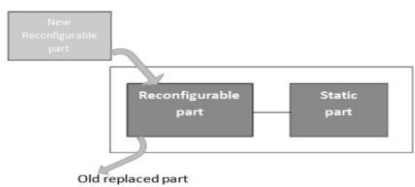

Figure 8: FPGA partial reconfiguration concept.

By using the FPGA PR feature, we enhance the multiplexing idea to provide more secure designs on FPGA. Each IP core will reserve a reconfigurable logic region, whereas the multiplexer or even the CRC voting circuit will reserve a static logic region. In order to decrease the chance of sending secret information, we suggest to partial reconfigure the FPGA to replace the current running IP with a new IP periodically. We will need a queue of IPs to let some of them run currently and others wait their turns in the queue.

Algorithm 1 describes the method of multiplexing reconfigurable IPs outputs for third party cores used on FPGA. The system designer keeps $m$ implementations of the suspected core from different vendors. The FPGA is configured only using 3 cores (for example), multiplexer, and CRC voting circuit. The system runs normally and the output is taken by multiplexing outputs of the different variants. Once the CRC voting circuit detects an error in an IP, its error count will be incremented. If it exceeds the chosen threshold, it will be marked as an infected IP and replaced with a new core. Anomalous core whose error count exceeds the threshold will be removed from the system using PR.

After a certain periodic time, we replace one of the three cores with the first IP on the waiting queue. Furthermore, the CRC results can be used to update the IP weights in the multiplexing process [24].

| **Algorithm 1:** Multiplexing reconfigurable IPs' outputs and CRC Trojan detection schema. ||
|---|---|
| **1:** | Download 3 variant cores; |
| **2:** | Run system normally; |
| **3:** | i = 0; |
| **4:** | **while** running normally **do** |
| **5:** |    **if** CRC error **then** |
| **6:** |       increment IP error counter; |
| **7:** |       **if** IP error counter exceeds threshold **then** |
| **8:** |          partial reconfigure FPGA to remove the infected IP totally; |
| **9:** |          replace the infected IP with the first IP on the queue; |
| **10:** |          mark the anomalous core as an infected core; |
| **11:** |       **end if** |
| **12:** |    **end if** |
| **13:** |    **if** periodic time elapsed **then** |
| **14:** |       use partial configuration to replace IP number *i* with the first IP on the queue; |
| **15:** |       i = (i+1) mod 3; |
| **16:** |    **end if** |
| **17:** | **end while** |

*B. Multiple Variant Implementation (MV)*

In this subsection, we show how to provide superior protection from hardware Trojans by using multiple variant implementations of suspected IPs. We propose adding multiple copies of the IP cores from different vendors and using the dynamic Trojan detector circuit to determine the suspected IP and the safe output using a majority voting circuit. Fig. 9 shows how the scheme works. We compare the output of *m* cores and if there is a mismatch between the different outputs, we suspect that there is a Trojan. If we have three or more IPs, we can use a voting circuit that chooses the correct output and raises a Trojan alarm, if needed [24].

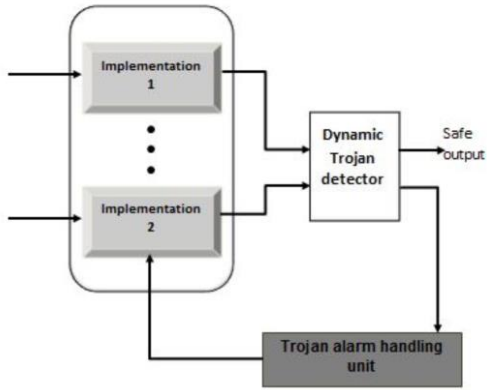

Figure 9: Using multiple variants of the same IP to catch Trojan.

The Trojan alarm handling unit can partially reconfigure the FPGA to remove the infected core and replace it with a new one while the rest of the FPGA logic is still running. The dynamic Trojan detector unit will also mark the anomalous core vendor as a suspected vendor.

The system designer keeps $m$ different implementations of the suspected core from different vendors. The FPGA is configured only using 3 cores (for example) and the Trojan Alarm circuit. The system runs normally and the output is taken by majority voting from the different variants. Once a Trojan alarm occurs, the anomalous core is declared. The core is replaced by another unsuspected core. We note that we will raise the Trojan alarm if IP error counter exceeds a suitable threshold to distinguish between a bug and a Trojan.

## 5. ATTACHING CRC LOGGER

In this section, we suggest to attach a CRC logger to the third party IP in the critical system, as shown in Fig. 10. CRC logger circuit must not come from a third party, as it might itself be infected. The CRC logger will basically compute the CRC for the IP inputs and outputs and store resulta in a memory, as shown in Fig. 11. Data stored in the memory can be extracted for comparison with actual IP specifications. The CRC logger can be enabled at critical times or randomly to track an IP behavior. We choose to us CRC outputs instead of actual inputs and outputs to decrease number of processed bits. CRC could capture 99.95% of transmission errors [25]. If a single bit is incorrect, the CRC value will not match up. We should highlight that data encoding can be used for further reduction in number of processed bits.

## 6. BUILDING A TRUSTED IP MARKET

As we have a lot of cores and vendors, we suggest establishing a core certificate authority, which will take the responsibility of handling the blacklist database. Any vendor who wants to have such a certificate has to contact the cores certificate authority. This authority will evaluate cores using the MV method. If a core is infected, the authority marks it as a spyware-infected core and the vendor has to prove to the authority that it is only a bug not a Trojan infection and can update it with a fixed version. Refer to Fig. 12 as an example on authority database. A core is labeled as safe if the warning score is zero, as buggy if the warning score is less than a threshold, and as infected otherwise. IP users can also give feedback about detected faults and thus indirectly share in updating database. This arrangement will help a designer to buy an IP from a vendor with a good reputation. This authority should be able to afford high cost hardware Trojan detection methodologies.

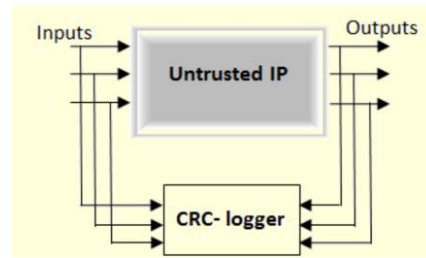

Figure 10: Attaching CRC logger to a third party IP to detect Trojan.

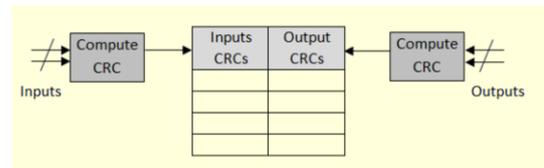

Figure 11: CRC-logger role: compute CRC and store result in memory.

| | Safe | Buggy | Infected |
|---|---|---|---|
| Core1 | ✓ | | |
| Core2 | | ✓ | |
| Core3 | | | ✓ |
| Core4 | | ✓ | |

Figure 12: A core certificate authority database example.

Since it is difficult that one user detects the Trojan on his/her own. So, our main idea is to have a centralized database to maintain the information and reinforce it. The certificate authority is responsible for maintaining the database and verifying the findings. However, users should collaborate and report discovered Trojans when a Trojan is caught. Each suspicious IP should be reported with the input vector at which the suspicious output occurred. The authority can also conduct its tests. However, a collaborative construction of the database is much more valuable due to the rare trigger of a Trojan.

We suggest certifying the safety of third party core usage by using the MV concept. We developed an automated system to build and update core certificate authority database for different cores and vendors, as shown in Fig. 13. Customers will submit their copies of the IP core from different vendors. Then, a Trojan detector circuit determines the suspected IP and the safe output using a majority voting among candidates. The circuit compares the output of *m* cores and whether there is a mismatch between the different outputs. Furthermore, it keeps track of the number of errors for the different cores. If such number exceeds the chosen threshold, a Trojan alarm will be raised to update the cores and vendor database. Infected IPs will be marked as Trojan-infected IP. Then, the server will partial reconfigure the FPGA to replace the Trojan-infected IP with a new one while the rest of the FPGA logic is still running. After a certain period, IP counters will be submitted to the server to determine safe and buggy IPs. Algorithm 2 describes the core authority automated system algorithm for third party cores. In our methodology each IP core reserves a reconfigurable logic region, whereas the Trojan detector unit reserves a static logic region.

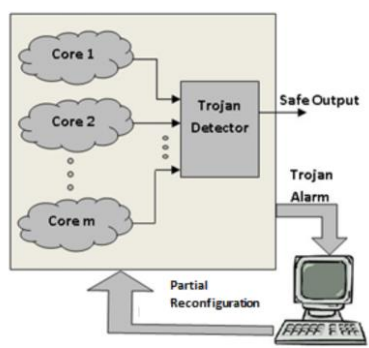

Figure 13: Automated system to build and update core certificate authority database.

| **Algorithm 2:** Core certificate authority automated system algorithm. ||
|---|---|
| **1:** | Download *m* variant cores; |
| **2:** | Run system normally; |
| **3:** | **while** time does not exceed time period **T do** |
| **4:** | compare IPs outputs and choose a major one using majority vote. |
| **5:** | **if** one IP output is different from others **then** |
| **6:** | increment IP error counter |
| **7:** | **if** IP error counter exceeds the threshold **then** |
| **8:** | raise Trojan alarm; |
| **9:** | use PR to replace infected core; |
| **10:** | mark the anomalous core as a Trojan-infected core in the database; |
| **11:** | end if |
| **12:** | end if |
| **13:** | end while |
| **14:** | Update server with current IPs error counters to update database with buggy and safe IPs |

## 7. ATTACKS AND COUNTERMEASURES

In this section, we describe some of the possible Trojan attacks and their countermeasures using our proposed methodologies.

### A. Internally-triggered Trojan to Leak Sensitive Information

The Trojan circuit in Fig. 14 sends the result of XORing the pseudo-random number generator (PRNG) output sequence with the secret information bits. The attacker knows about the exact setup of the code generator, which is required to predict the code sequence. So, the attacker can easily XOR the transmitted data again to get the secret key. Lin and Kasper introduced Trojan for ASIC, which sends information wireless through a separate leakage circuit [26]. However, we assume Trojan for an FPGA with a simple modification by sending the Trojan output multiplexed with current encrypted data output in the same clock encryption period or simply using unused bits in network packets. We classify this Trojan as an internally triggered Trojan.

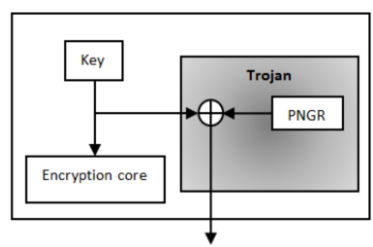

Figure 14: Hardware Trojan to leak secret key.

Proposed SB can deal perfectly with the internally-triggered Trojan as sensitive information will be well-encrypted. The MV method will choose the majority output and thus an infected IP can be detected. On the other hand, multiplexing IPs' outputs will try to decrease the probability of losing information. However, an attacker can predict his/her IP turn due to using pseudo-random generator and sends the information bits distributed on multiple turns. CRC enhancement will try to solve this issue by detecting the infected IP and removing it. However, some parts of the sensitive information maybe lost before Trojan detection.

The main advantage of the SB method is that it works regardless of the source of the trigger. The method proposed by Waksman and Sethumadhavan in [13] cannot deal with the internally-triggered Trojan as it depends on preventing external triggers. Moreover, they added special hardware for each Trojan activation method so more methods will need more hardware.

If the third party IP is within a network and the Trojan aims to leak network information, we suggest using SB to obfuscate data of all third party IP output lines to prevent data leakage and then do deobfuscation at trusted IPs inputs, as shown in Fig. 15.

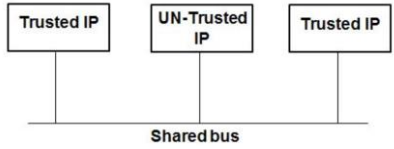

Figure 15: Suspected third party IP within a network.

### B. Externally-triggered Trojan to Leak Sensitive Information

The discussion in this subsection is similar to the case of an internally-triggered Trojan as we do not depend on preventing triggering the Trojan. Waksman and Sethumadhavan in [13] can only deal with a specific set of Trojan triggers. However, our methods work for any trigger type as we do not try to stop the trigger, but rather try to stop the impact of an activated Trojan.

If we analyze more the MCRC schema, we find that there is a period (T) before the detection of the infected IP. System output during this period is shared between all used IPs. If we assume a fair IP multiplexing, we will find that the attacker IP will have a (T/number of IPs) period to leak sensitive information. On the other hand, the MV schema will not allow any period to the attacker, as it will choose the major IPs' output during the running time. We may remark the former argument as an advantage for MV over MCRC. It will provide higher level of protection when addressing leaking information and disturbing functionality Trojans. Nevertheless, we find that MCRC can get rid of injected Trojan. So, we must highlight that as an advantage for MCRC over MRVO.

### C. Trojan to Disrupt Functionality

SB cannot help with disrupting functionality Trojan. MRVO will decrease the Trojan effect only. Interestingly, the higher the number of variants (*m*), the lower the Trojan effect. But, unfortunately higher *m* imposes higher cost. MCRC and MV will detect the infected Trojan and remove it from the system perfectly. However, CRC enhancement may result in function disruption before Trojan detection.

### D. Timing and Power Side Channel Attacks

Timing and power side channel attacks can happen with SB. However, other methodologies can resist these attacks as the system contains multiple implementations and other supplementary circuits and therefore will absorb both timing and power side channels. We summarize these information in Table I. Each methodology has a score against each attack. A score of 0 means the method cannot protect against the attack. Scores of 1, 2, and 3 refer to low, moderate, and strong protection against the attack, respectively. Method overall score is obtained by summing the attack scores, then scaling to be out of 3. It can be seen that MCRC and MV provide the highest two overall scores and are expected to introduce the largest overheads (as will be show in Section 8).

TABLE I. TROJANS AND PROPOSED METHODS EVALUATION AGAINST ATTACKS.

| Trojan type | SB | MRVO | MCRC | MV |
|---|---|---|---|---|
| Externally-triggered | 3 | 1 | 2 | 3 |
| Internally-triggered | 3 | 1 | 2 | 3 |
| Leaking information | 3 | 1 | 2 | 3 |
| Disturbing functionality | 0 | 1 | 2 | 3 |
| Power/timing side channel | 0 | 3 | 3 | 3 |
| **Method overall score** | **1.8** | **1.4** | **2.2** | **3** |

## 8. EXPERIMENTAL RESULTS

In this section, we study the effect of using the proposed methods on consumed power, number of used look-up tables (LUTs), and critical path delay. We will describe in details the used benchmarks and study the overhead of each method. First, we will study the overheads of using protection from Trojan. Then, we will concentrate on Trojan detection schemes.

### A. Protection Against Trojan

In this subsection, we study the power, area, and delay overheads of using SB and MRVO.

#### 1) SB Overhead

The six OpenCores benchmarks listed in Table II are used to study the SB method [27].

TABLE II. DESCRIPTION OF USED BENCHMARKS (BENCHMARKS ARE SELECTED FROM OPENCORES [27]).

| Benchmark | Description |
|---|---|
| **FHT** | Fast Hadamhard Transform (FHT) of 8-bit input data using matrix addition. |
| **PID** | Digital proportional–integral–derivative (PID) controller. |
| **F3M** | $Y^3$ mod $(x^{97} + x^{12} + 2)$, where Y is a 194-bit input port. |
| **FPU** | A set of floating point operations. |
| **CA PRNG** | 1-D binary cellular automata with wrap around at the edges, i.e., a ring. |
| **Solomon** | Reed Solomon Decoder with a 204-byte and 188-byte input and output codeword lengths, respectively. |

We implemented the selected six benchmarks on Xilinx Virtex 6 FPGAs and used Xilinx ISE Design Suite v14 1 to evaluate the experimental results. We used an xc6vcx75tff484-1 FPGA device for all the benchmarks, except the FPU one for which we used an xc6vcx130tff1156-1 device, as it needs more FPGA resources. Also, power is calculated using a 100 MHz clock, except for FPU for which a 10 MHz clock rate was used due to the 99.495 ns minimum allowed clock period.

After running synthesis, we got the number of used LUTs (from the synthesis XST report), the critical path delay (from the synthesis timing report), and consumed power (using Xilinx Xpower Analyzer). Summary of results obtained before adding the SB method is shown in Table III.

TABLE III.  DETAILS FROM THE HARDWARE IMPLMENTATION OF THE BENCHMARKS IN TABLE II BEFORE ADDING THE SB METHOD.

| Benchmark | Used LUTs | Delay (ns) | Consumed power (mW) |
|---|---|---|---|
| FHT | 26 | 1.720 | 1.300 |
| PID | 1345 | 3.938 | 1.321 |
| F3M | 144 | 1.297 | 2.350 |
| FPU | 9591 | 99.495 | 1.327 |
| CA PRNG | 105 | 1.824 | 1.343 |
| Solomon | 4954 | 4.479 | 1.386 |

We used the obfuscating function described in Subsection 3.A to protect benchmarks outputs from a Trojan embedded in a third party IP. Table IV summarizes the results after integrating the obfuscating function. If we compare Table III and Table IV, we find an average increase of 12.74% in used LUTs. We also find using our obfuscating function with F3M introduces more delay than other benchmarks as the size of its output port is 194 bits, so it needs more processing time. Finally, we conclude that the average increase in critical path delay is 2.309% and the average increase in consumed power is 0.766% in our benchmarks.

TABLE IV.  SIMPLE OBFUSCATION FUNCTION RESULTS.

| Benchmark | Percentage of increase in Used LUTs | Percentage of increase in delay | Percentage of increase in consumer power |
|---|---|---|---|
| FHT | 0.46 | 0.520 | 0.846 |
| PID | 3.79 | 0.001 | 0.529 |
| F3M | 39.58 | 12.950 | 1.957 |
| FPU | 0.84 | 0.001 | 0.075 |
| CA PRNG | 31.43 | 0.380 | 1.116 |
| Solomon | 0.34 | 0.001 | 0.072 |
| Average | 12.74 | 2.309 | 0.766 |

*2) MRVO Overhead*

In this subsection, we study the effect of the MRVO method on the number of used LUTs, power, and PR time and evaluate the third party authority system cost.

ZedBoard Zynq Evaluation and Development Kit (xc7z020clg484-1) was used in this experiment as it supports PR. In experiments requiring multiple implementations of the same IP, we use an ALU (logic_unit) core and a serial transmitter unit (RS_TX) core. For each core, we get multiple implementations from different vendors to study the overhead of using multiple cores of used IPs.

We selected two different implementations of each IP and integrated them with the multiplexing circuit to construct the full system. Table V summarizes implementation details of the used RS_TX IPs without using PR. We note here that the reported consumed power does not include the ZedBoard leakage power.

TABLE V.  DETAILS FROM THE HARDWARE IMPLMENTATION OF THE USED RS_TX CORES WITHOUT USING PR.

|  | Used LUTs | Consumed power (mW) |
|---|---|---|
| RS_TX_n1 | 18 | 0.70 |
| RS_TX_n2 | 24 | 0.78 |
| RS_TX_n3 | 30 | 0.85 |

We investigate the effect of using the PR on the consumed power and the number of used LUTs. Table VI shows the percentage increase in number of LUTs and the percentage increase in the consumed power when we set a reconfigurable area for each IP. *RS_TX_1PR_n1* indicates using one reconfigurable area only for *RS_TX_n1*. The table shows that using a reconfigurable area increases the required LUTs. If we compare results of both Table V and Table VI, we will observe the overhead of using PR on a single programmed IP only. The average increase in number of LUTs is 31.53% and the average increase in consumed power is 1.2733%.

TABLE VI.  DETAILS FROM THE HARDWARE IMPLMENTATION OF THE USED RS_TX CORES WHEN USING PR.

|  | Percentage of increase in Used LUTs | Percentage of increase in consumer power |
|---|---|---|
| RS_TX_1PR_n1 | 55.56 | 1.2500 |
| RS_TX_1PR_n2 | 27.27 | 1.3800 |
| RS_TX_1PR_n3 | 11.76 | 1.1900 |
| Average | 31.53 | 1.2733 |

Then, we investigate the effect of multiplexing two IPs without PR. Table VII illustrates the variant number of LUTs and consumed power if we use every two IPs in the same design with multiplexer and no PR. *RS_TX-T_n1n2* indicates the combination of *RS_TX_n1* and *RS_TX_n2*. If we compare results in Table V and Table VII, we see an increase in the allocated area which is expected due to loading two designs on the FPGA. This increase is proportional to the size of the two selected IPs.

TABLE VII.  DETAILS FROM THE HARDWARE IMPLMENTATION OF MULTIPLEXING TWO RS_TX CORES WITHOUT USING PR.

|  | Used LUTs | Consumed power (mW) |
|---|---|---|
| RS_TX_n1n2 | 48 | 1.22 |
| RS_TX_n1n3 | 53 | 1.25 |
| RS_TX_n2n3 | 57 | 1.35 |

Furthermore, we investigate the effect of multiplexing two RS_TX IPs in case of reserving reconfigurable area for each IP. Table VIII shows the number of LUTs, consumed power, and switching time when we set two reconfigurable areas in which IPs can be replaced by each other. *RS_TX_2PR_n1n2* indicates using *RS_TX_n1* and *RS_TX_n2* at the same time, each in an individual reconfigurable area. Results in Table VIII indicate that using two reconfigurable areas increases the required LUTs. Moreover, if two empty black box modules (RS_TX-2PR_BB) are loaded, we will also need a number of LUTs which is larger than what is required by one reconfigurable area (12 LUTs and 1.01 mW for one reconfigurable area). If we compare results in Table VII and Table VIII, we find that using two reconfigurable areas with PR need more LUTs than mixing two IPs without PR. However, in PR these two modules can be easily replaced by *m* modules.

TABLE VIII. DETAILS FROM THE HARDWARE IMPLMENTATION OF MULTIPLEXING TWO RS_TX CORES WHEN USING PR.

|  | Used LUTs | Consumed power (mW) |
|---|---|---|
| **RS_TX-2PR_n1n2** | 65 | 1.94 |
| **RS_TX-2PR_n1n3** | 64 | 1.83 |
| **RS_TX-2PR_n2n3** | 75 | 1.96 |
| **RS_TX-2PR_BB** | 25 | 1.26 |

Since PR switches between IPs, we found that the time elapsed to replace one design with another ranges from 400 to 500 msec, with an average of 450 msec. While the time used to program the whole design ranges from 5 to 6 sec. As a result, we notice that PR succeeds in replacing designs at a negligible time compared to the time required to program the whole design.

We repeat the former experiments, but using ALU IPs from different vendors. Table IX reports the number of LUTs and consumed power if we use every IP alone. In the table, *logic_unit_n1* indicates the first ALU IP.

TABLE IX. DETAILS FROM THE HARDWARE IMPLMENTATION OF THE USED ALU CORES WITHOUT USING PR.

|  | Used LUTs | Consumed power (mW) |
|---|---|---|
| **logic_unit_n1** | 7 | 0.15 |
| **logic_unit_n2** | 9 | 0.21 |
| **logic_unit_n3** | 4 | 0.12 |

Table X shows the number of LUTs and consumed power in case of multiplexing two ALUs without PR. *logic_unit_n1n2* indicates combination of the first and the second implementations of the logic unit core (*logic_unit_n1* and *logic_unit_n2*).

TABLE X. DETAILS FROM THE HARDWARE IMPLMENTATION OF MULTIPLEXING TWO ALU CORES WITHOUT USING PR.

|  | Used LUTs | Consumed power (mW) |
|---|---|---|
| **logic_unit_n1n2** | 15 | 0.28 |
| **logic_unit_n1n3** | 14 | 0.19 |
| **logic_unit_n2n3** | 18 | 0.27 |

Finally, Table XI reports the number of LUTs, consumed power, and switching time when we reserve reconfigurable area for each ALU IP.

TABLE XI. DETAILS FROM THE HARDWARE IMPLMENTATION OF MULTIPLEXING TWO ALU CORES WHEN USING PR.

|  | Used LUTs | Consumed power (mW) |
|---|---|---|
| **logic_unit-2PR_n1n2** | 43 | 0.45 |
| **logic_unit-2PR_n1n3** | 38 | 0.36 |
| **logic_unit-2PR_n2n3** | 40 | 0.42 |

*3) Different Selection Criteria*

In this subsection, we show the effect of using biased random selection of IPs on the number of used LUTs, critical path delay, and its accuracy.

ZedBoard Zynq Evaluation and Development Kit (xc7z020clg484-1) was also used in this experiment. We used three different implementations of the RS_TX core to select from them.

We study the percentage of selecting an infected IP using unbiased and biased selection. In our experiment, we use two non-infected IPs and an infected one.

In one experiment, the Trojan in the infected IP is active during all clock cycles. In this experiment, if an infected IP is selected, the output will be infected. In another one, the Trojan is active at the odd cycles only. In this one, if an infected IP is selected, output may or may not be infected.

In the biased selection, we give an 8-bit weight to each IP. Initially, all IPs have the same weight, which is half the maximum weight. While running, we increment weights of IPs with same output by one and decrement weights of odd IPs by one. We are limited in incrementing and decrementing by the number of bits of used weights.

We ran the system for 1000 clock cycles and recorded the probability of selecting an infected IP and an infected output in Table XII. The table shows how the accuracy of selecting a correct IP and a correct output is improved in the biased selection.

TABLE XII. PROBABILITY OF SELECTING AN INFECTED IP AND AN INFECTED OUTPUT USING BIASED AND UNBIASED SELECTION.

|  | Probability of selecting an infected IP | Probability of selecting an infected output |
|---|---|---|
| *Trojan of infected IP is active in all cycles* | | |
| **Unbiased selection** | 30% | 30% |
| **Biased selection** | 2% | 2% |
| *Trojan of infected IP is active in odd cycles* | | |
| **Unbiased selection** | 30% | 15% |
| **Biased selection** | 20% | 10% |

Next, we show the overhead of using the biased selection. Table XIII shows the number of LUTs and critical path delay for the biased and unbiased selection. From the table, we conclude that using biased selection has an obvious overhead on number of LUTs, but it does not affect the critical path delay.

TABLE XIII. DETAILS FROM THE HARDWARE IMPLMENTATION OF THE BIASED AND UNBIASED SELECTION.

|  | Used LUTs | Delay (ns) |
|---|---|---|
| **Unbiased selection** | 57 | 3.273 |
| **Biased selection** | 149 | 3.273 |

*B. Trojan Detection Schemes*

In this subsection, we study the overhead of the different Trojan detection schemes: MCRC and MV.

*1) MCRC Overhead*

We used three UART IPs as multiple variants. CRC computation is done on each UART output. We note that using CRC in this case may not be considered as a type of overhead as UART is usually used with CRC computation to detect sending errors. CRC truncated polynomial is $x^5+x^2+1$ and it is applied on each eight bits of the UART output. Table XIV summarizes the synthesis details using a Virtex 6 xc6vcx75tff484-1 device.

TABLE XIV. DETAILS FROM THE HARDWARE IMPLMENTATION OF THE MCRC METHOD ON THE UART CORES.

|  | Used LUTs | Delay (ns) | Consumed power (mW) |
|---|---|---|---|
| **UART1** | 26 | 1.874 | 13 |
| **UART2** | 28 | 1.657 | 6 |
| **UART3** | 68 | 2.371 | 10 |
| **FinalUART** | 179 | 2.651 | 107 |

We repeat the experiment with the ALU IPs. Table XV summarizes the synthesis details using a Virtex 6 xc6vcx75tff484-1 device. Leakage power equals 1.293 mW and it is the same for all cases.

TABLE XV. DETAILS FROM THE HARDWARE IMPLMENTATION OF THE MCRC METHOD ON THE ALU CORES.

|  | Used LUTs | Delay (ns) | Consumed power (mW) |
|---|---|---|---|
| **ALU1** | 48 | 12.220 | 2 |
| **ALU2** | 36 | 4.623 | 13 |
| **ALU3** | 169 | 3.767 | 50 |
| **FinalALU** | 221 | 12.528 | 101 |

*2) MV Overhead*

We used three UART IPs as multiple variants and integrated them with the dynamic Trojan detector circuit to construct FinalUART. Table XVI shows the synthesis details using a Virtex 6 xc6vcx75tff484-1 device.

We notice that the FinalUART delay equals the maximum delay of a UART core plus the delay of dynamic Trojan detector circuit. However, the latter delay in this experiment is less than 1 ps due to the low processing time of a 1-bit output. Leakage power equals 1.293 mW and it is the same for all cases.

TABLE XVI. DETAILS FROM THE HARDWARE IMPLMENTATION OF THE MV METHOD ON THE UART CORES.

|  | Used LUTs | Delay (ns) | Consumed power (mW) |
|---|---|---|---|
| **UART1** | 26 | 1.874 | 13 |
| **UART2** | 28 | 1.657 | 6 |
| **UART3** | 68 | 2.371 | 10 |
| **FinalUART** | 123 | 2.371 | 22 |

We repeat the former experiment, but using ALU IPs from different vendors. Table XVII shows the synthesis details using a Virtex 6 xc6vcx75tff484-1 device.

We notice that the FinalALU delay equals the maximum delay of an ALU core plus the delay of dynamic Trojan detector circuit. The latter delay in this experiment equals 1.695 ns. Leakage power equals 1.293 mW and it is the same for all cases.

TABLE XVII. DETAILS FROM THE HARDWARE IMPLMENTATION OF THE MV METHOD ON THE ALU CORES.

|  | Used LUTs | Delay (ns) | Consumed power (mW) |
|---|---|---|---|
| **ALU1** | 48 | 12.220 | 2 |
| **ALU2** | 36 | 4.623 | 13 |
| **ALU3** | 169 | 3.767 | 50 |
| **FinalALU** | 190 | 13.915 | 100 |

We repeat the same experiment, but using AES IPs from different vendors. The AES core is a symmetric block cipher core that can process data blocks of 128 bits, using a 128-bit key. Table XVIII shows the synthesis details using a Virtex 6 xc6vlx550tff1760-1 device.

We notice that the FinalAES delay equals the maximum delay of an AES core plus the delay of dynamic Trojan detector circuit. The latter delay in this experiment equals 0.214 ns. Leakage power equals 3.769 mW and it is the same for all cases.

TABLE XVIII. AES MULTIPLE VARIANT METHODOLOGIES RESULTS.

|  | Used LUTs | Delay (ns) | Consumed power (mW) |
|---|---|---|---|
| **AES1** | 702 | 4.313 | 461 |
| **AES2** | 1553 | 5.502 | 130 |
| **AES3** | 256 | 2.972 | 419 |
| **FinalAES** | 1153 | 5.716 | 1050 |

## 9. CONCLUSION AND FUTURE WORK

We have introduced the first methodologies that work by embracing Trojans and protecting the system from them without trying to detect the Trojan during testing. Our methods operate at runtime instead of the traditional test-time techniques. This work targets protection from Trojans that could be embedded in third party IPs, where no golden design is available. We presented methods for simple system level protection and more sophisticated methods that are capable of detecting the Trojan and removing it from the system. The SB methodology was the first simple protection scheme introduced and worked by obfuscating the output of suspected IPs to hide any information leakage. The second simple protection method was the MRVO method that protects the system from both leaking sensitive information and malfunction. We also introduced two methods: MCRC and MV to dynamically remove Trojan-infected IPs and report Trojans. They take advantage of the PR feature of modern FPGAs to clean up infected IPs.

We provided sample implementations of the different methodologies. In the SB methodology, the delay overhead was negligible, whereas the power overhead did not exceed 2%. Other methodologies provide higher overhead. However they provide superior security mechanisms. To the best of our knowledge, this is the first work that attempts to do dynamic hardware Trojan detection and protection on FPGAs.

We are looking forward to analyzing our protection and detection techniques on ASICs instead of FPGAs. Proposing hardware Trojans and attacking different systems should give a chance for better comparison between proposed schemas. We think about injecting hardware Trojans in one of IPs variants and compare MV and MCRC schemes. Comparing the strength of the proposed techniques from the point of view of probabilistic study of defense should be interesting also. We will try to put guidelines to determine the strength of any protection or detection technique.

## REFERENCES

[1] M. Banga, M. S. Hsiao, "Vitamin: Voltage inversion technique to ascertain malicious insertions in ICs," In Proceedings of the IEEE International Workshop on Hardware-Oriented Security and Trust (HOST '09), Washington, DC, USA, 2009, pp. 104–107.

[2] R. Rad, J. Plusquellic, M. Tehranipoor, "A sensitivity analysis of power signal methods for detecting hardware Trojans under real process and environmental conditions," IEEE Transactions on Very Large Scale Integrated Systems (VLSI) (2010), 18(12), pp. 1735–1744.

[3] J. Rajendran, V. Jyothi, O. Sinanoglu, R. Karri, "Design and analysis of ring oscillator based design-for-trust technique," In Proceedings of the IEEE 29th VLSI Test Symposium (VTS), Dana Point, CA, USA, 2011, pp. 105–110.

[4] H. Salmani, M. Tehranipoor, J. Plusquellic, "New design strategy for improving hardware Trojan detection and reducing Trojan activation time," In Proceedings of the IEEE International Workshop on Hardware-Oriented Security and Trust (HOST '09), Washington, DC, USA, 2009, pp. 66–73.

[5] M. Banga, M. Chandrasekar, L. Fang, M. Hsiao, "Guided test generation for isolation and detection of embedded Trojans in ICs," In Proceedings of the 18th ACM Great Lakes symposium on VLSI (GLSVLSI '08), New York, NY, USA, 2008, pp. 363–366.

[6] M. Banga, M. S. Hsiao, "A novel sustained vector technique for the detection of hardware Trojans," In Proceedings of the 22nd International Conference on VLSI Design (VLSID '09), Washington, DC, USA, 2009, pp. 327–332.

[7] D. Du, S. Narasimhan, R. Chakraborty, S. Bhunia, "Self-referencing: A scalable side-channel approach for hardware Trojan detection," In Proceedings of the Cryptographic Hardware and Embedded Systems (CHES 2010), Santa Barbara, CA, USA, 2010, pp. 173–187.

[8] Y. Jin, Y. Makris, "Hardware Trojan detection using path delay fingerprint," In Proceedings of the IEEE International Workshop on Hardware-Oriented Security and Trust (HOST '09), Washington, DC, USA, 2008, pp. 51–57.

[9] F. Koushanfar, A. Mirhoseini, Y. Alkabani, "A unified submodular framework for multimodal IC Trojan detection," In Proceedings of the 12th international conference on Information hiding (IH'10), Calgary, AB, Canada, 2010, pp. 17–32.

[10] M. Potkonjak, A. Nahapetian, M. Nelson, T. Massey, "Hardware Trojan horse detection using gate-level characterization," In Proceedings of the 46th Annual Design Automation Conference (DAC '09), New York, NY, USA, 2009, pp. 688–693.

[11] X. Zhang, M. Tehranipoor, "Case study: Detecting hardware Trojans in third-party digital IP cores," In Proceedings of the IEEE International Workshop on Hardware-Oriented Security and Trust (HOST '11), San Diego, CA, USA, 2011, pp. 67–70.

[12] M. Potkonjak, "Synthesis of trustable ics using untrusted CAD tools," In Proceedings of the 2010 47th ACM/IEEE Design Automation Conference (DAC), 2010, pp. 633–634.

[13] A. Waksman, S. Sethumadhavan, "Silencing hardware backdoors," In Proceedings of the 2011 IEEE Symposium on Security and Privacy (SP '11), Berkeley/Oakland, CA, USA, 2011, pp. 49–63.

[14] M. Beaumont, B. Hopkins, T. Newby, "Safer path: Security architecture using fragmented execution and replication for protection against Trojaned hardware," In Proceedings of the IEEE Conference on Design, Automation and Test in Europe (DATE '12), Dresden, Germany, 2012, pp. 1000–1005.

[15] A. Baumgarten, A. Tyagi, J. Zambreno, "Preventing IC piracy using reconfigurable logic barriers," IEEE Design & Test of Computers (2010), 27(1), pp. 66–75.

[16] M. Tarek Ibn Ziad, A. Al-Anwar, Y. Alkabani, M. W. El-Kharashi, H. Bedour, "E-voting attacks and countermeasures," In Proceedings of the 10th International Symposium on Frontiers of Information Systems and Network Aplications (FINA 2014), held in conjunction with the 28th IEEE International Conference on Advanced Information Networking and Applications (AINA-2014), Victoria, BC, Canada, 2014, pp. 269–274.

[17] M. Tehranipoor, F. Koushanfar, "A survey of hardware Trojan taxonomy and detection," IEEE Design & Test of Computers (2010), 27(1), pp. 10–25.

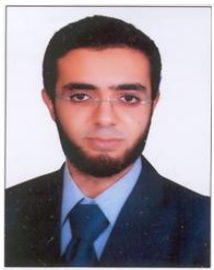

**Amr Alanwar** received the B.Sc. degree (first class honors) and the M.Sc. degree in computer engineering from Ain Shams University, Cairo, Egypt, in 2010 and 2013, respectively. He is currently a Ph.D. candidate in the Networked & Embedded Systems Laboratory (NESL) at the University of California, Los Angeles (UCLA), Los Angeles, CA, USA. His M.Sc. thesis was on the protection and detection of hardware Trojans. He worked as a hardware R&D Engineer in Emulation Division (MED) at Mentor Graphics and then joined the L-1 Identity Solutions corporate research center at Safran Morpho. He was also a research and teaching assistant in the Department of Computer and Systems Engineering, Ain Shams University, Cairo, Egypt during the period December 2011—December 2013. His research interests are security, control, image processing, and robotics.

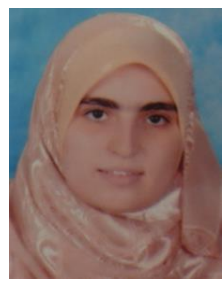

**Mona A. Aboelnaga** received the B.Sc. degree (first class honors) in computer engineering from Ain Shams University, Cairo, Egypt, in 2013. She is currently an M.Sc. student and a teaching assistant in the Department of Computer and Systems Engineering, Ain Shams University, Cairo, Egypt. Her research interests include cyber-physical systems security and machine learning.

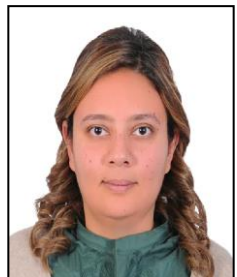

**Yousra Alkabani** holds B.Sc. and M.Sc.degrees in computer and systems engineering from Ain Shams University in 2003 and 2006, respectively. She received a Ph.D. in Computer Science from Rice University in December 2010. She is an Assistant Professor of Computer and Systems Engineering at Ain Shams University since May 2011 and a visiting Assistant Professor of Computer Science and Engineering at the American University in Cairo (AUC) since 2013. Her research interests include hardware security, low power design, and embedded systems.

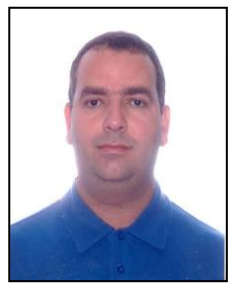

**M. Watheq El-Kharashi** received the Ph.D. degree in computer engineering from the University of Victoria, Victoria, BC, Canada, in 2002, and the B.Sc. degree (first class honors) and the M.Sc. degree in computer engineering from Ain Shams University, Cairo, Egypt, in 1992 and 1996, respectively. He is a Professor in the Department of Computer and Systems Engineering, Ain Shams University, Cairo, Egypt and an Adjunct Professor in the Department of Electrical and Computer Engineering, University of Victoria, Victoria, BC, Canada. His general research interests are in advanced system architectures, especially networks-on-chip (NoC), systems-on-chip (SoC), and secure hardware. He published more than 100 papers in refereed international journals and conferences and authored two books and 6 book chapters.

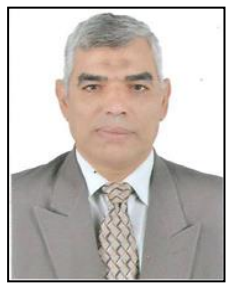

**Hassan Bedour** received the Ph.D. degree in computer engineering from Ain Shams University, Cairo, Egypt in 1990. He is currently an Associate Professor in the Department of Computer and Systems Engineering, Ain Shams University, Cairo, Egypt. His research interests are in semantic computing, intelligent systems, computer architecture, and system design. He held several ICT managerial positions and participated in national ICT projects as a consultant and in several international research projects.